\documentclass{aa}
\usepackage{longtable}
\usepackage{supertabular}
\usepackage{graphics}
\usepackage{rotating}
\usepackage{natbib}
\begin{document}
\title{RASS-SDSS Galaxy Cluster Survey.}  
\subtitle{VI. The dependence of the cluster SFR on the cluster global properties.}
\author{P. Popesso\inst{1}, A. Biviano\inst{2}, M. Romaniello\inst{1}, 
H. B\"ohringer\inst{3}} 
\institute{ European Southern Observatory,
Karl Schwarzschild Strasse 2, D-85748 \and INAF - Osservatorio
Astronomico di Trieste, via G. B. Tiepolo 11, I-34131, Trieste, Italy
\and Max-Planck-Institut fur extraterrestrische Physik, 85748
Garching, Germany}

\abstract
{To quantify the relationships between
star formation in cluster galaxies and global cluster properties.
Using a subsample of 79 nearby clusters from the RASS-SDSS galaxy
cluster catalogue of Popesso et al. (2005a), we perform a regression
analysis between the cluster integrated star formation rate ($\Sigma
SFR$) the cluster total stellar mass ($M_{\star}$), the fractions of
star forming ($f_{SF}$) and blue ($f_b$) galaxies and other cluster
global properties, namely its richness ($N_{gal}$, i.e. the total
number of cluster members within the cluster virial radius, corrected
for incompleteness), velocity dispersion ($\sigma_v$), virial mass
($M_{200}$), and X-ray luminosity ($L_X$).  All cluster global
quantities are corrected for projection effects before the
analysis. Galaxy $SFR$s and stellar masses are taken from the catalog
of Brinchmann et al. (2004), which is based on SDSS spectra. We only
consider galaxies with $M_r \leq -20.25$ in our analysis, and exclude
AGNs. We find that both $\Sigma SFR$ and $M_{\star}$ are
correlated with all the cluster global quantities. A partial
correlation analysis show that all the correlations are induced by the
fundamental one between $\Sigma SFR$ and $N_{gal}$, hence there is no
evidence that the cluster properties affect the mean $SFR$ or
$M_{\star}$ per galaxy. The relations between $\Sigma SFR$ and
$M_{\star}$, on one side, and both $N_{gal}$ and $M_{200}$, on the
other side, are linear, i.e. we see no evidence that different
clusters have different $SFR$ or different $M_{\star}$ per galaxy and
per unit mass. The fraction $f_{SF}$ does not depend on any cluster
property considered, while $f_b$ does depend on $L_X$. We note that a
significant fraction of star-forming cluster galaxies are red
($\sim$ 25\% of the whole cluster galaxy population). We
conclude that the global cluster properties are unable to affect the
$SF$ properties of cluster galaxies, but the presence of the X-ray
luminous intra-cluster medium can affect their colors, perhaps
through the ram-pressure stripping mechanism.}


\maketitle
\authorrunning{P. Popesso et al.}
\titlerunning{RASS-SDSS Galaxy Cluster Survey. VI.}

\section{Introduction}
What role does the environment play in the evolution of cluster
galaxies? The dependence of the morphological mix from the
environmental conditions was qualitatively illustrated in the early
study of the Virgo cluster by Hubble and Humason (1931) and has been
confirmed in many studies (e.g. Oemler 1974; Dressler 1980; Postman \&
Geller 1984; Dressler et al. 1997). The clear observational evidence
is that the high density regions, such as the massive galaxy clusters,
are dominated by a quiescent early type galaxy population, while the
late type star forming galaxies more likely populate low density
regions such as the field. A recently proposed way to study the
relation between galaxy population and environmental conditions is the
analysis of the ongoing star formation (SF) in galaxies of different
environments (see, e.g., Christlein \& Zabludoff 2005).  The SF rate
(SFR) is an important measure of the evolutionary state of a galaxy,
and a sensitive indicator of the environmental interactions. Previous
studies of cluster galaxy SFRs have sometimes reached conflicting
conclusions. The SFRs of cluster galaxies have been found to be
reduced (Kennicutt 1983; Bicay and Giovanelli 1987; Kodaira et
al. 1990; Moss \& Whittle 1993; Abraham et al. 1996; Balogh et
l. 1998, 2002; Koopmann \& Kenney 1998; Hashimoto et al. 1998; Gavazzi
et al. 2002; Pimbblet et al. 2006), comparable (Kennicutt et
al. 1984; Donas et al. 1990; Gavazzi et al. 1991, 1998; Biviano et
al. 1997; Moss \& Whittle 2005), or in some case enhanced (Moss \&
Whittle 1993; Bennet \& Moss 1998) relative to the SFRs of field
galaxies of the same classes.

Several cluster-related environmental processes can affect the SFRs of
galaxies.  Some processes mainly affect the gaseous content of a
galaxy, such as the ram-pressure stripping (Gunn \& Gott 1972; Kenney
et al. 2004; van Gorkom 2004), re-accretion of the stripped gas
(Vollmer et al. 2001), turbulence and viscosity (e.q. Quilis et al. 2001),
and starvation/strangulation (Larson et al. 1980). Gravitational
processes, which affect both the gaseous and the stellar properties of
a galaxy, range from low-velocity tidal interactions and mergers
(e.g. Mamon 1996; Barnes \& Hernquist 1996; Conselice 2006), to
high-velocity interactions between galaxies and/or clusters (Moore et
al. 1998, 1999; Struck 1999; Mihos 2004). Despite a number of recent
studies of nearby and distant clusters, it is not yet clear which of
these processes, if any, are dominant. 

Clues on the relative importance of the cluster-related environmental
processes can be obtained by investigating the evolution of the
star-forming properties of cluster galaxies. In this context, the most
important evolutionary phenomenon is the Butcher-Oemler (BO hereafter)
effect (Butcher \& Oemler 1978, 1984), i.e. the increasing fraction of
blue cluster members with redshift. The BO effect has been confirmed
and detailed by many studies since the original works of Butcher \&
Oemler (e.g.  Ellingson et al. 2001; Margoniner et al. 2001; Alexov et
al. 2003; De Propris et al. 2003; Rakos and Shombert 2005),
although Andreon et al. (2004, 2006) have argued that no
cluster-dependent evolution is required to explain the BO effect,
which is entirely compatible with the normal color evolution of
galaxies in an ageing universe. The BO-effect is purely
photometrical. The spectroscopic version of the BO-effect is an excess
of emission-line and star-forming galaxies in distant, relative to
nearby, clusters, first suggested by Dressler \& Gunn (1982) and later
confirmed by several authors (e.g. Postman et al. 1998, 2001; Dressler
et al. 1999; Finn et al. 2004, 2005; Homeier et al. 2005; Poggianti et
al. 2006, P06 hereafter)

Most of the analyses so far have concentrated on the comparison of the
star-forming properties of individual cluster galaxies with those of
field galaxies, and on the variation of the galaxy SFRs on the local
density of their environment. However, it is also important to assess
the dependence (if any) of the star-forming properties of cluster
galaxies on their cluster global properties, such as the mass,
velocity dispersion and X-ray luminosity. Should the SFRs of cluster
galaxies depend on global properties of their host cluster, results
obtained for different individual clusters would not be
straightforward to compare, thereby producing apparently discrepant
results. Moreover, the relative efficiency of the different
evolutionary processes depends on several cluster properties, and
investigating the SFRs of cluster galaxies as a function of these
properties can help understanding this issue (see, e.g., Pimbblet
2003). Also the evolution of the star forming properties of cluster
galaxies must be studied in close connection with the evolution of
their host cluster properties. In fact, evolutionary studies of
cluster galaxy SFRs may be affected by selection biases if the SFRs
depend on global cluster properties, such as their masses. Since in
flux-limited surveys more massive clusters are preferentially selected
with increasing redshift, a biased estimate of the evolution of the
star-forming properties of cluster galaxies may result (see, e.g.,
Newberry et al. 1998; Andreon \& Ettori 1999),

Recently, several studies have addressed the dependence of the
star-forming properties of cluster galaxies on their host global
properties. Several studies have found that the cluster global
properties do not affect the star-forming properties of cluster
galaxies. In particular, no dependence has been found of either the
blue or the late-type galaxy fraction in clusters on cluster velocity
dispersions ($\sigma_v$s) and masses (Goto 2005), nor of the blue
fraction with cluster richness, concentration, and degree of
subclustering (De Propris et al. 2004). On the other hand, both
Margoniner et al. (2001) and Goto et al. (2003) had previously found a
dependence of the blue or late-type galaxy fractions on the cluster
richness. Goto (2005) has also claimed no dependence on the cluster
$\sigma_v$s and masses of either the total cluster SFR or of the total
cluster SFR normalized by the cluster mass, in disagreement with Finn
et al. (2005) who have shown that the integrated SFR per cluster mass
decreases with increasing cluster mass. Lea \& Henry (1988), Fairley
et al. (2002), and Wake et al. (2005) have all failed to find any
dependence of the fraction of blue cluster galaxies with the cluster
X-ray luminosity, $L_X$.  Similarly, Balogh et al. (2002) have
compared the galaxy SFRs in high-$L_X$ and low-$L_X$ clusters and have
found no differences. In the sample of Homeier et al. (2005) there are
hints of correlations between the total cluster SFRs and cluster
$L_X$s and intra-cluster gas temperatures, $T_X$s, but the trends are
not really significant.  Most recently, P06 have
found that the fraction of emission-line galaxies (ELGs hereafter)
decreases with increasing cluster $\sigma_v$. The trend is continuous
at high-$z$, but is characterized by a break at $\sigma_v \sim
500$--600 km~s$^{-1}$ in nearby clusters, where the relation they find
is consistent with the results obtained by Biviano et al. (1997).

In this paper we re-address the issue of the dependence of the SFR and
the fraction of star forming galaxies on the cluster global
properties. At variance with most previous studies, we consider both
optical and X-ray cluster global properties, namely the mass,
$\sigma_v$, and $L_X$. While these quantities are correlated (Popesso
et al. 2005a, Paper III of this series), it is worthwhile to consider them all,
since the star-forming properties of cluster galaxies may show a
stronger dependence on one of these properties, thereby pointing to a
different physical mechanism affecting their SFRs. E.g., Postman et
al. (2005) have recently shown that the fraction of early-type
galaxies in distant clusters does depend on $L_X$, but not on
$\sigma_v$, nor on $T_X$.  In our analysis we use a sample of 79
low-redshift clusters taken from the X-ray selected RASS-SDSS galaxy
cluster catalog (Popesso et al. 2004, Paper I) and the optically selected Abell
cluster sample (Popesso et al. 2006a, Paper V). Besides providing further
constraints on the mechanisms of galaxy evolution in clusters, our
investigation should be useful for assessing the possible selection
effects in the comparison of the star-forming properties of galaxies
in nearby vs. distant clusters, as well as in clusters at similar
redshifts but with different global properties.

In Sect.~\ref{s-data} of the paper we describe our dataset.  In
Sect.~\ref{s-sfr} we analyze the relation between the cluster integrated
star formation rate and the global properties of the systems. In
Sect.~\ref{s-frac} we apply the same analysis to the fraction of blue
cluster galaxies and the fraction of cluster star forming
galaxies. Sect.~\ref{s-conc} contains our conclusions.

Throughout this paper, we use $H_0=70$ km s$^{-1}$ Mpc$^{-1}$ in a
flat cosmology with $\Omega_0=0.3$ and $\Omega_{\Lambda}=0.7$
(e.g. Tegmark et al. 2004).

\section{The data}
\label{s-data}
The optical data used in this paper are taken from the Sloan Digital
Sky Survey (SDSS, Fukugita et al. 1996, Gunn et al. 1998, Lupton et al. 1999,
York et al. 2000, Hogg et al. 2001, Eisenstein et al. 2001, Smith et
al. 2002, Strauss et al. 2002, Stoughton et al.  2002, Blanton et
al. 2003 and Abazajian et al.  2003).  The SDSS consists of an imaging
survey of $\pi$ steradians of the northern sky in the five passbands
$u, g, r ,i, z,$ in the entire optical range.  The imaging survey
is taken in drift-scan mode.  The imaging data are processed with a
photometric pipeline (PHOTO, Lupton et al. 2001) specially written for
the SDSS data.  For each cluster we defined a photometric galaxy
catalog as described in Section 3 of Paper~I (see also
Yasuda et al. 2001).  For the analysis in this paper we use only SDSS
Model magnitudes.

The spectroscopic component of the survey is carried out using two
fiber-fed double spectrographs, covering the wavelength range
3800--9200 \AA, over 4098 pixels. They have a resolution
$\Delta\lambda/\lambda$ varying between 1850 and 2200, and together
they are fed by 640 fibers, each with an entrance diameter of 3
arcsec. The fibers are manually plugged into plates inserted into the
focal plane; the mapping of fibers to plates is carried out by a
tiling algorithm (Blanton et al. 2003) that optimizes observing
efficiency in the presence of large-scale structure. 

The X-ray data are taken from the ROSAT All Sky Survey. The RASS was
conducted mainly during the first half year of the ROSAT mission in
1990 and 1991 (Tr\"umper 1988). The ROSAT mirror system and the
Position Sensitive Proportional counter (PSPC) operating in the soft
X-ray regime (0.1-2.4 keV) provided optimal conditions for the studies
of celestial objects with low surface brightness. In particular, due
to the unlimited field of view of the RASS and the low background of
the PSPC, the properties of nearby clusters of galaxies can be ideally
investigated.

\subsection{The cluster sample}
In this paper we use a combined sample of X-ray selected galaxy
clusters and optically selected systems.  The X-ray selected clusters
are taken from the RASS-SDSS galaxy cluster catalog of paper III, and
the optically selected clusters are taken from the sample of Abell
clusters spectroscopically confirmed using SDSS DR3 data of Paper
V. Of these clusters, we only consider those with available X-ray
center, in order to minimize possible centering errors. There is
partial overlap between the X-ray and optical samples. In Paper V we
have recently compared the properties and scaling relations of
optically- and X-ray selected clusters. We have found no difference
among the two samples, except for a larger scatter of the $L_X$-mass
relation when derived on the optically-selected clusters rather than
on the X-ray selected ones (see Paper V for details). We can thus
safely combine the two samples together in the present analysis.

We have determined the cluster membership by studying the redshifts
distribution of the galaxies in the cluster region (see next section
for details). In order to analyze the SFR and the blue fraction of
galaxies in the same magnitude range for all the clusters, we have
selected only those clusters for which the limiting magnitude of the
SDSS spectroscopic catalog, $r_{Petro} \le 17.77$, corresponds to an
an absolute magnitude limit fainter than $-20.25$ (and hence to a
redshift limit $z \sim 0.1$). This magnitude is about 0.7 mag fainter
than the value of $M^{\star}$ of the Schechter (1976) function that
provides the best-fit to the RASS-SDSS clusters luminosity function
(Popesso et al. 2006b, Paper IV). Among these clusters, we finally
select only those containing at least 5 cluster members brighter than
$-20.25$ in the $r$-band. Note that the $\sigma_v$s and masses of
these clusters are estimated using all cluster members, irrespectively
of their magnitude, and hence are generally based on at least 10
cluster members. Studying clusters extracted from cosmological
simulations, Biviano et al. (2006) have recently shown that 10 cluster
members are sufficient to obtain an unbiased estimate of a cluster
$\sigma_v$.  The final catalog contains 79 clusters, spanning a large
mass range ($10^{13}$--$5\times10^{15} M{\odot}$).

\subsection{Cluster masses, velocity dispersions and X-ray luminosities}
We here provide a summary of the methods by which we measure the
cluster global properties, $\sigma_v$s, masses, and $L_X$s.  Full
details can be found in paper III and IV.

We define the cluster membership of a galaxy on the basis of its
location in the projected phase-space diagram, velocity with respect
to the cluster mean vs. clustercentric distance. Specifically, we
combine the methods of Girardi et al. (1993) and Katgert et
al. (2004). Using the cluster members, the virial analysis (see, e.g.,
Girardi et al. 1998) is then performed on the clusters with at least
10 member galaxies. The line-of-sight velocity dispersion is computed
in the cluster rest-frame (Harrison 1974) using the biweight
estimator (Beers et al. 1990). By multiplying it by a factor
$\sqrt{3}$ we obtain the 3D $\sigma_v$.  The virial masses, $M_{200}$
are corrected for the surface pressure term (The \& White 1986) and
estimated at the virial radius, $r_{200}$, using an iterative
procedure.  Namely, we start by using Carlberg et al.'s
1997 $r_{200}$ definition as a first guess, then extrapolate or
interpolate the virial mass estimate obtained within the observational
aperture to $r_{200}$ using a Navarro et al. (1997) mass
profile. This mass estimate is used to obtain a new estimate of
$r_{200}$ and the virial mass is finally re-estimated by extrapolating
or interpolating the observed value to the new estimate of $r_{200}$
(see Biviano et al. 2006 for a thorough description of our
procedure).

Cluster $L_X$s are calculated from RASS data using the growth curve
analysis method (B\"ohringer et al. 2000).  This method is optimized
for the detection of the extended emission of clusters by assessing
the plateau of the background subtracted cumulative count rate
curve. The X-ray luminosity estimate we adopt corresponds to the total
flux inside the radius $r_{200}$, corrected for the missing flux by
using a standard $\beta$-model for the X-ray surface brightness (see
B\"ohringer et al. 2000 for more details). The correction is typically
only $8 - 10\%$.

\subsection{Galaxy Star Formation Rates}
We take the SFR-estimates for our cluster members from Brinchmann et
al. (2004, hereafter, B04). In addition to SFRs, we also take from
B04 the SFRs normalized to the stellar masses,
$SFR/m^*$.  They provide mainly $H_{\alpha}$-derived SFR, based
on SDSS spectra, for all the SDSS DR2 spectroscopic catalog. B04
divided their galaxy sample in three subsamples on the basis of the
Baldwin et al. (1981) $\log [OIII]5007/H \beta$
vs. $\log [NII]6584/H \alpha$ diagram. B04 distinguish the following
galaxy categories: star-forming galaxies, composite galaxies, AGNs,
and unclassifiable objects. For all the star-forming galaxies and the
unclassifiable objects the SFR is calculated directly from the
emission lines (see B04 for details).

B04 provide three estimators of the galaxy SFR, the median, the mode
and the average of the likelihood distribution. Since the average and
the mode of the distribution are somewhat binning sensitive, we adopt
the median of the distribution as our SFR estimator.  B04's SFRs are
derived from spectra taken within the 3 arcsec diameter fibers of the
SDSS, which generally sample only a fraction of the total galaxy
light. B04 correct their SFRs for these aperture effects (see B04 for
details), and we adopt their corrected (total) SFRs. We have checked
that our results do not change when instead of the median we use the
mode, and when instead of the corrected SFRs we use the uncorrected
ones.

\begin{figure}
\begin{center}
\begin{minipage}{0.49\textwidth}
\resizebox{\hsize}{!}{\includegraphics{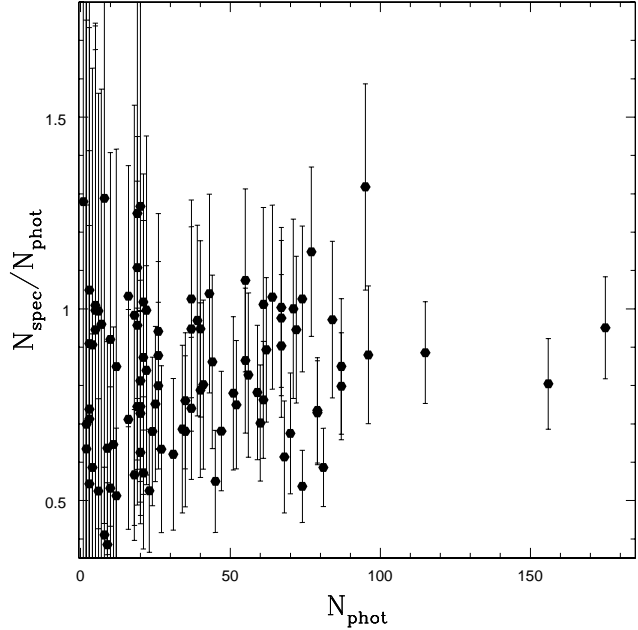}}
\end{minipage}
\end{center}
\caption{ Comparison between the number of cluster spectroscopic
members ($N_{spec}$) within $r_{200}$ and with $r_{petro}\le -20.25$
with the number of cluster photometric members ($N_{phot}$) in the
same region and magnitude range. The inverse of the
$N_{spec}/N_{phot}$ ratio gives the incompleteness correction factor
to apply to the $\Sigma SFR$. When this factor is lower than 1, we set
it to 1.}
\label{norma}
\end{figure}

\begin{figure}
\begin{center}
\begin{minipage}{0.49\textwidth}
\resizebox{\hsize}{!}{\includegraphics{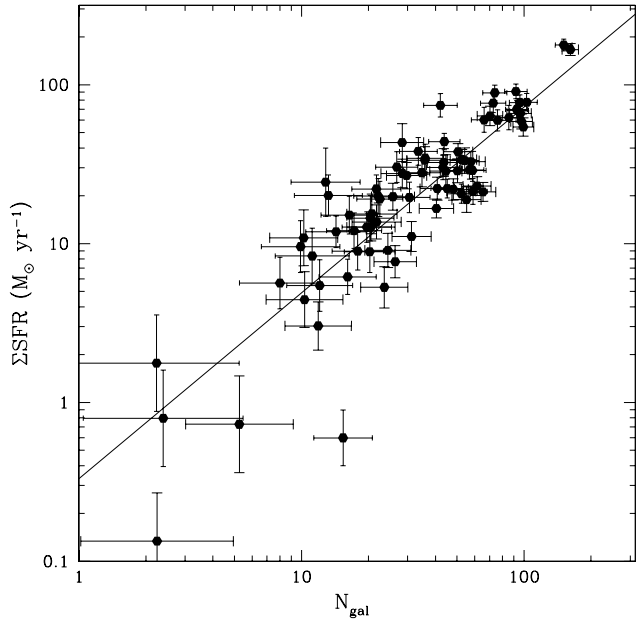}}
\end{minipage}
\end{center}
\caption{
Correlation of the integrated cluster SFR calculated within $r_{200}$
and with $r_{petro} \le -20.25$ with the total number of galaxies in the
same region and magnitude range. We define $N_{gal}$ by subtracting
statistically the background and foreground galaxies.}
\label{ngal}
\end{figure}

\begin{figure*}
\begin{center}
\begin{minipage}{0.32\textwidth}
\resizebox{\hsize}{!}{\includegraphics{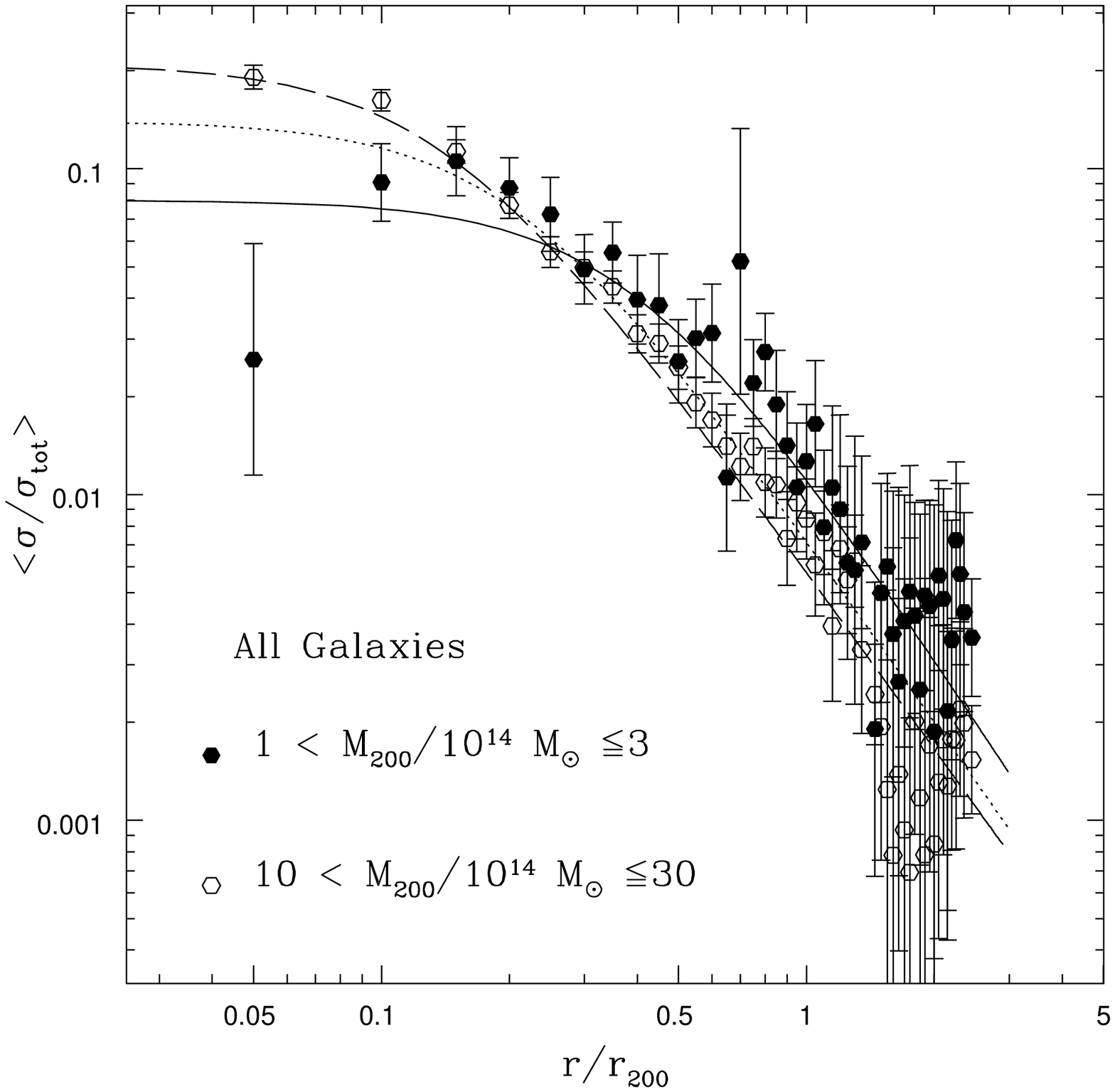}}
\end{minipage}
\begin{minipage}{0.32\textwidth}
\resizebox{\hsize}{!}{\includegraphics{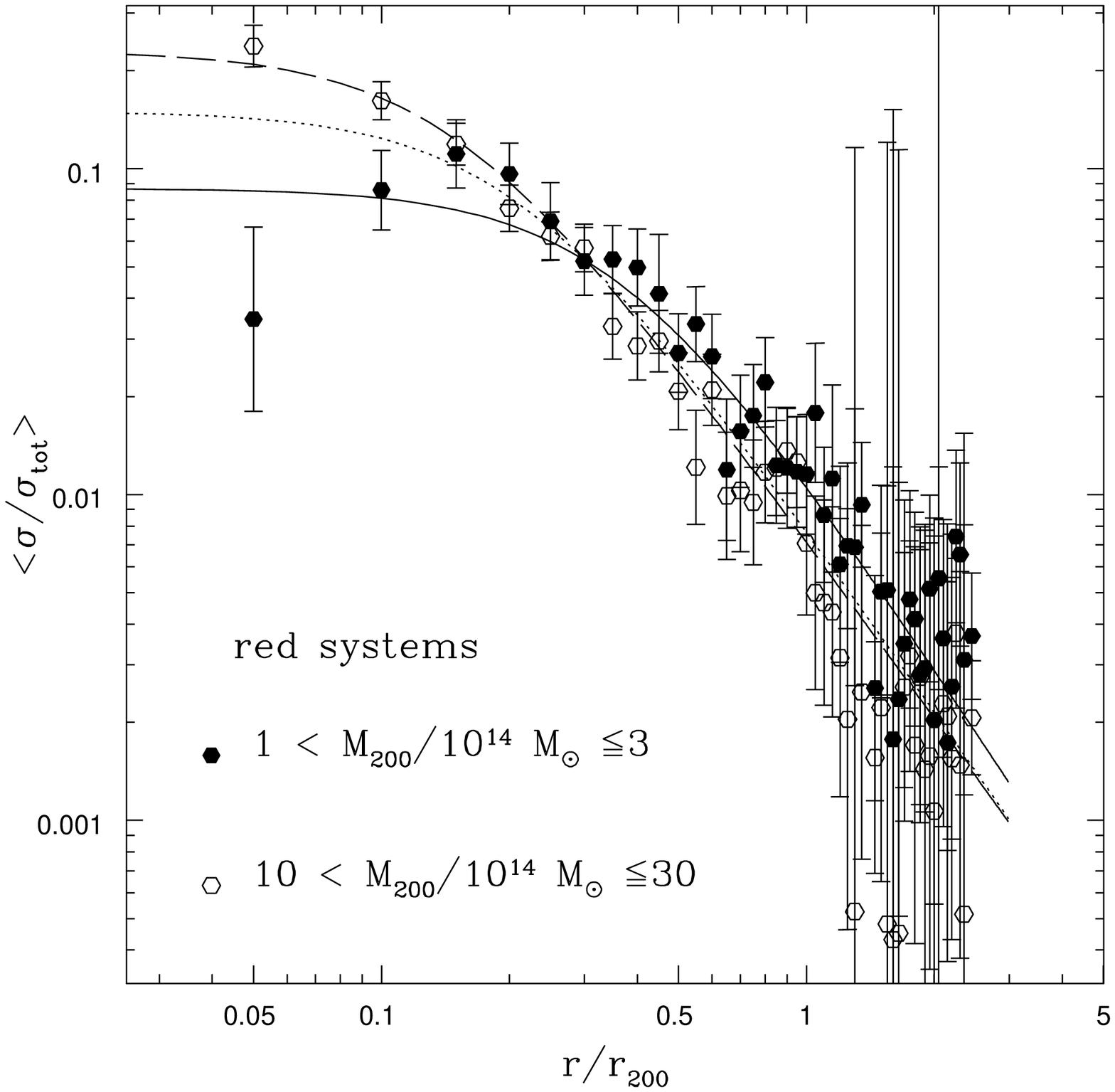}}
\end{minipage}
\begin{minipage}{0.32\textwidth}
\resizebox{\hsize}{!}{\includegraphics{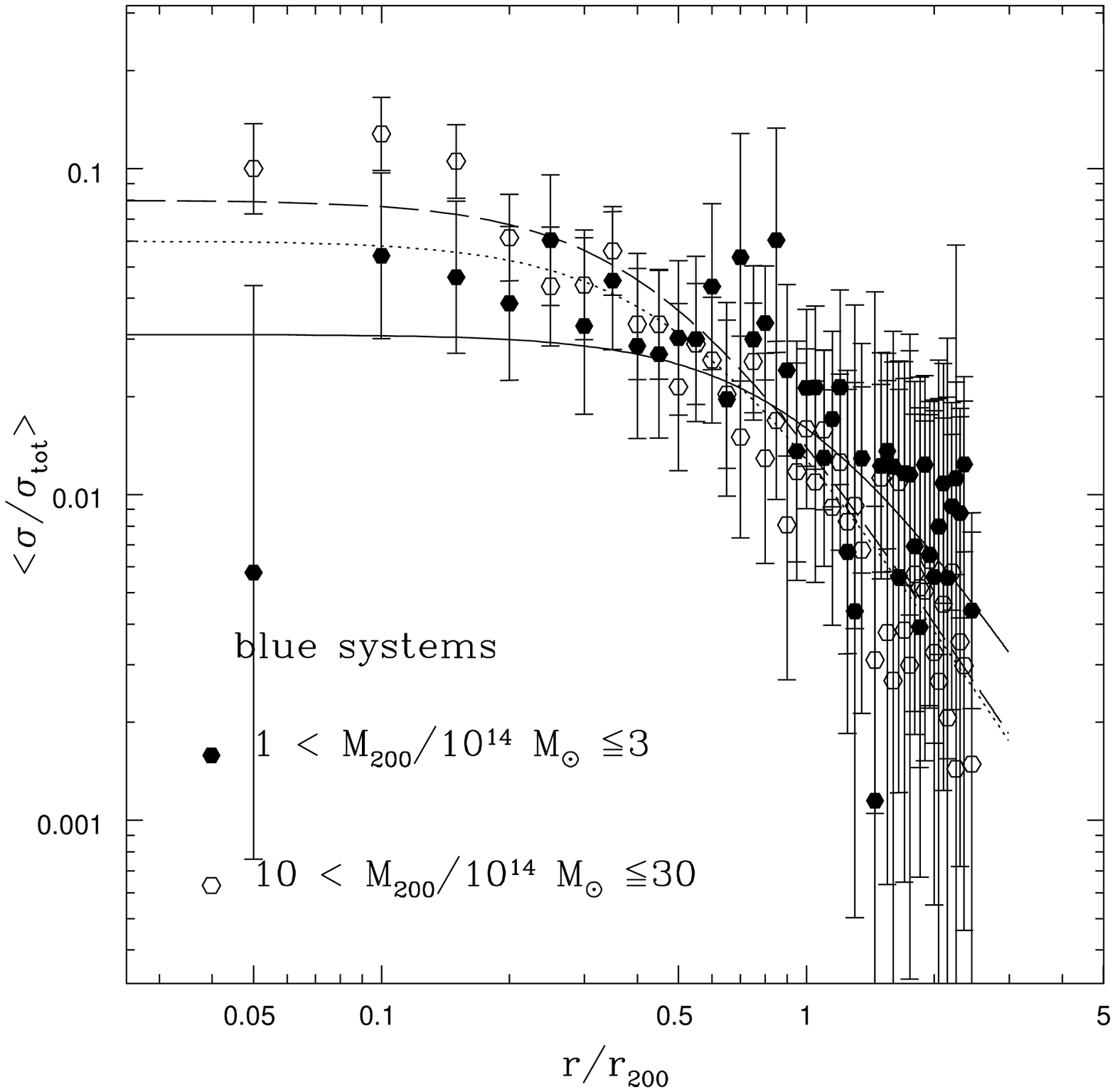}}
\end{minipage}
\end{center}
\caption{
The stacked surface number density profiles of clusters in lowest and
highest mass bins for the whole (left panel), red (central panel) and
the blue (right panel) cluster galaxy population. The individual
cluster profiles are obtained by considering all the galaxies with
$r_{petro} < -18.5$.  In each panel the open circles are the density
profile of the highest mass bin and the filled circles are the profile
of the lowest mass bin. The dashed line is the best-fit King profile of
the highest mass bin, the dotted line is the best-fit King profile of the
mean surface number density distribution (obtained stacking all the
clusters in the sample) and the solid line is the best-fit King profile of
the lowest mass bin. }
\label{plotto}
\end{figure*}

\subsection{Cluster Star Formation Rates}
\label{csfr}
In order to estimate the integrated cluster SFRs we first sum up the
SFRs of their cluster members, AGNs and composite-spectrum galaxies
excluded. I.e. we consider all the galaxies classified star-forming by
B04, as well as the unclassifiable objects. The unclassifiable objects
among our cluster members have extremely low SFR (as estimated by B04)
and their summed contribution to the cluster integrated SFR is not
significant.

Since our spectroscopic sample is not complete down to the chosen
magnitude limit, we need to multiply the sum of the cluster member
SFRs by an incompleteness correction factor. In order to estimate the
incompleteness correction for each cluster we compare the number of
cluster spectroscopic members, $N_{spec}$, within $r_{200}$ and with
$r_{petro} \le -20.25$, with the corresponding number of cluster
galaxies estimated from the photometric data, $N_{phot}$, since the
photometric sample is complete for $r_{petro} \le -20.25$. In order to
estimate $N_{phot}$ we first estimate the density of foreground and
background galaxies from the counts of $r_{petro} \le -20.25$ galaxies
in an annulus outside the virialized area (at radii $> r_{200}$)
centered on the cluster center. We then subtract the number of
background galaxies expected in the cluster area from the number of
galaxies (down to the same magnitude limit) in the cluster region. In
Fig. \ref{norma} we show the number ratios of spectroscopic and
photometric members as a function of $N_{phot}$. 80\% of our clusters
have a completeness level higher than 80\%. We calculate the
incompleteness correction factor as the maximum between
$N_{phot}/N_{spec}$ and 1.

Another correction we need to apply to the sum of cluster member SFRs
is the de-projection correction since the global cluster quantities we
want to compare the integrated SFR with, are all de-projected
quantities.  When we sum up the SFRs of cluster members with a
clustercentric projected distance $\leq r_{200}$, we include the
contribution of galaxies outside the virial sphere, but within the
cylinder of same radius. In Fig. \ref{ngal} we shows the relation
between the integrated SFR within $r_{200}$ and $N_{spec}$. Because of
the strict proportionality between these two quantities, and because
the relation is linear within the errors (see Table \ref{table1}), we
can estimate the de-projection correction for the number of cluster
members, and apply the same correction to the integrated SFR. In order
to estimate the de-projection correction for $N_{spec}$, we build the
number density profiles of our clusters, and fit them with the King
(1962) cored profile, and the NFW cuspy profile (Navarro et
al. 1997). We then de-project these profiles, and take the ratio
between the integrals from the center to $r_{200}$ of the de-projected
and the projected profiles. This ratio provides the correction factor.

The number density profiles of our clusters are built by stacking
together our clusters after rescaling their galaxy clustercentric
distances by their cluster $r_{200}$s (see also Popesso et al. 2006c,
Paper VII, where we perform the same analysis). We use the SDSS
$r$-band photometric data down to the completeness limit $r=21$, and
consider a common absolute magnitude limit of $-18.5$ for all our
clusters. The cluster galaxy distributions are normalized to the total
number of galaxies within $r_{200}$, after subtraction of the mean
background galaxy density, evaluated within the $2.5-3.5 \times
r_{200}$ annulus.  We split our sample of clusters in 6 mass bins
($M_{200}/10^{14} M_{\odot} \le 1$, $1 < M_{200}/10^{14} M_{\odot} \le
3$, $3 < M_{200}/10^{14} M_{\odot} \le 7$, $7 < M_{200}/10^{14}
M_{\odot} \le 10$,$10 < M_{200}/10^{14} M_{\odot} \le 30$, and
$M_{200}/10^{14} M_{\odot} > 30$) and determine the number density
profile for each of these subsamples.  Each bin contains at least 10
clusters. We find that the number density profiles become steeper near
the center as the cluster mass increases. This is true independently
for the red and blue cluster members ($u-r \geq 2.22$ and,
respectively, $< 2.22$, see Strateva et al. 2001), so this is not an
effect due to the population of cluster galaxies, but it is a
mass-related effect.  More massive clusters have more centrally
concentrated galaxy distributions. The best fit parameters of the King
profiles for different cluster mass bins and galaxy populations are
listed in Table \ref{table2}.  In Fig. \ref{plotto} we show the number
density profiles in the lowest and highest mass bins for the whole
(left panel), the red (central panel) and the blue (right panel)
cluster galaxy populations.

Since the galaxy number density profiles depend on the mass of the
cluster, also the de-projection corrections are mass dependent. In
Table \ref{table2} we list the correction factors determined for each
mass bin by using the best-fit King profiles for the whole cluster
population. We apply these mass-dependent de-projection correction
factors to the integrated SFRs. In the following, $\Sigma SFR$ refers
to the incompleteness- and de-projection-corrected values of the
integrated SFRs within a sphere of radius $r_{200}$.

\begin{table*}
\caption{King's profile best fit parameters for
different cluster mass bins ($m=M_{200}/(10^{14} M_{\odot})$) and
different cluster galaxy population.
$r_c$ is the core radius of the King profile
expressed in units of $r_{200}$. $\sigma_0$ is the central number
density of galaxy normalized to the total number of
galaxies. 'cf' is the de-projection correction factor to be
applied to the observed number of cluster members within a
projected clustercentric distance $r_{200}$.}
\begin{center}
\begin{tabular}[b]{ccccccc}
\hline
\renewcommand{\arraystretch}{0.2}\renewcommand{\tabcolsep}{0.05cm} 
 & $m \le 1$ & $1 < m \le 3$ & $3 < m \le 7$  & $7 < m \le 10$ & $10 < m \le 30$  & $m > 30$  \\ \hline
\multicolumn{7}{c}{The whole cluster galaxy population} \\ \hline
$r_c$ & 0.40$\pm$0.08 & $0.20\pm0.02$ & $0.22\pm0.01$ & $0.16\pm0.01$ & $0.15\pm0.01$ & $0.15\pm0.01$  \\
$\sigma_0$ & $0.08\pm0.01$ & $0.12\pm0.01$ & $0.14\pm0.01$ & $0.20\pm0.01$ & $0.19\pm0.01$ & $0.19\pm0.01$ \\
cf & 0.72 & 0.79 & 0.81 & 0.81 & 0.85 & 0.85 \\ 
\hline
\multicolumn{7}{c}{The red galaxy population (u-r $>$ 2.22)} \\ \hline
$r_c$ & 0.37$\pm$0.03 & $0.23\pm0.01$ & $0.20\pm0.01$ & $0.17\pm0.01$ & $0.17\pm0.01$ & $0.15\pm0.01$  \\
$\sigma_0$ & $0.08\pm0.01$ & $0.13\pm0.01$ & $0.16\pm0.01$ & $0.19\pm0.01$ & $0.20\pm0.01$ & $0.23\pm0.01$ \\
\hline
\multicolumn{7}{c}{The blue galaxy population (u-r $<$ 2.22)} \\ \hline
$r_c$ & 1.03$\pm$0.13 & $0.65\pm0.06$ & $0.46\pm0.05$ & $0.36\pm0.08$ & $0.34\pm0.04$ & $0.46\pm0.06$  \\
$\sigma_0$ & $0.03\pm0.01$ & $0.05\pm0.01$ & $0.07\pm0.01$ & $0.08\pm0.01$ & $0.09\pm0.01$ & $0.08\pm0.01$ \\
\hline
\end{tabular}
\label{table2}
\end{center}							   
\end{table*}

\begin{table*}
\caption{Best-fit parameters of the correlations
between global cluster quantities, $A$ vs. $B$, with
$A=10^{\beta} \times B^{\alpha}$, and the estimate of
the orthogonal scatter,
expressed in dex. Errors on the best-fit parameters are given at the 95\%
confidence level. Units are:
$M_{\odot} \, yr^{-1}$ for $\Sigma SFR$,$M_{\odot}$ for $M_{200}$ and
$M_{\star}$, km~s$^{-1}$ for $\sigma_v$ and $10^{44} \rm{erg~s}^{-1}$ for
$L_X$ ($f_b$ is unit-less).}
\begin{center}
\begin{tabular}[b]{cc|ccc|cc}
\renewcommand{\arraystretch}{0.2}\renewcommand{\tabcolsep}{0.05cm}
A& B& $\alpha $ & $\beta$ & $\sigma$ & $r_S$ & $P(r_S)$ \\ \hline
$\Sigma SFR$ & $N_{gal}$                &  1.08 $\pm$  0.08&   -0.32  $\pm$ 0.12&  0.13 &  0.84 & $2\times 10^{-21}$\\
$\Sigma SFR$ & $M_{200}$                &  1.11 $\pm$  0.10&  -15.36  $\pm$ 1.57&  0.20 &  0.74 & $1\times 10^{-16}$\\
$\Sigma SFR$ & $\sigma_v$               &  2.18 $\pm$  0.23&   -4.96  $\pm$ 0.62&  0.15 &  0.76 & $2\times 10^{-18}$\\
$\Sigma SFR$ & $L_X$                    &  0.62 $\pm$  0.09&    1.62  $\pm$ 0.07&  0.27 &  0.46 & $2\times 10^{-5}$\\
$\Sigma SFR/M_{200}$ & $\sigma_v$       & -0.67 $\pm$  0.19&  -11.63  $\pm$ 0.54&  0.24 & -0.30 & $4\times 10^{-3}$ \\
$M_{\star}$    &$\Sigma SFR$             &  1.09 $\pm$  0.06&  -11.77  $\pm$ 0.73&  0.12 &  0.80 & $2\times 10^{-17}$\\
$M_{\star}$ & $N_{gal}$                  &  1.01 $\pm$  0.07&   10.86  $\pm$ 0.07&  0.07 &  0.85 & $2\times 10^{-21}$\\
$M_{\star}$ & $M_{200}$                  &  1.08 $\pm$  0.09&   -3.50  $\pm$ 1.12&  0.16 &  0.75 & $4\times 10^{-16}$ \\
$M_{\star}$ & $\sigma_v$                 &  2.31 $\pm$  0.23&    5.55  $\pm$ 0.63&  0.09 &  0.84 & $3\times 10^{-20}$ \\
$M_{\star}$ & $L_X$                      &  0.61 $\pm$  0.06&   12.47  $\pm$ 0.04&  0.25 &  0.49 & $2\times 10^{-5}$ \\
$f_b$ & $L_X$                           & -0.13 $\pm$  0.04&    0.91  $\pm$ 0.04&  0.19 & -0.41 & $2\times 10^{-4}$\\
$\Sigma SFR_{blue}/\Sigma SFR$ & $L_X$  & -0.19 $\pm$  0.05&    0.78  $\pm$ 0.04&  0.22 & -0.41 & $4\times 10^{-4}$\\
\hline
\end{tabular}
\label{table1}
\end{center}							   
\end{table*}

\section{The dependence of the cluster $\Sigma SFR$ 
on the cluster global properties.}
\label{s-sfr}
\begin{figure}[h]
\begin{center}
\begin{minipage}{0.49\textwidth}
\resizebox{\hsize}{!}{\includegraphics{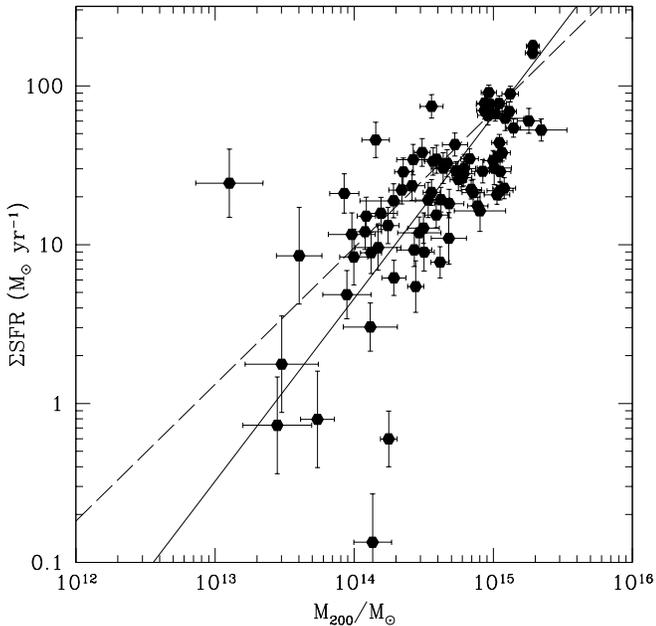}}
\end{minipage}
\end{center}
\caption{
$\Sigma SFR$ vs. $M_{200}$. The
solid line is the best fit obtained using the de-projected quantities.
The dashed line is the best fit we would obtain without correcting
the integrated cluster SFR for the projection effects.}
\label{SFR}
\end{figure}

In order to analyse the relation between $\Sigma SFR$ and $M_{200}$ we
perform an orthogonal linear regression in the logarithmic space, using
the software package ODRPACK (Akritas $\&$ Bershady 1996). We find a
significant correlation between these two quantities (as quantified by
the Spearman correlation coefficient, see Table \ref{table1}). The
slope of the relation is consistent with unity (see Table
\ref{table1}, ). Fig. \ref{SFR} shows the $\Sigma SFR-M_{200}$
relation. Note that the slope of the relation would have been found to
be significantly smaller than unity, had we not applied the
de-projection correction to $\Sigma SFR$.

\begin{figure}
\begin{center}
\begin{minipage}{0.49\textwidth}
\resizebox{\hsize}{!}{\includegraphics{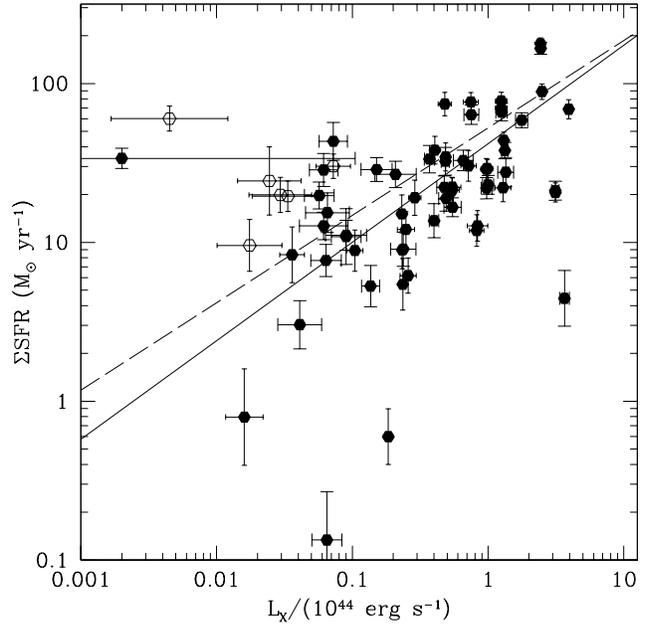}}
\end{minipage}
\end{center}
\caption{ $\Sigma SFR$ vs. $L_X$ relation. Open points are the
Abell X-ray-Underluminous (AXU) clusters (for details, see paper V of
this series).  The solid line is the best fit obtained after
correction for projection effects. The dashed line is the best fit we
would obtain had we not applied the de-projection correction.}
\label{lx}
\end{figure}

\begin{figure}[h]
\begin{center}
\begin{minipage}{0.49\textwidth}
\resizebox{\hsize}{!}{\includegraphics{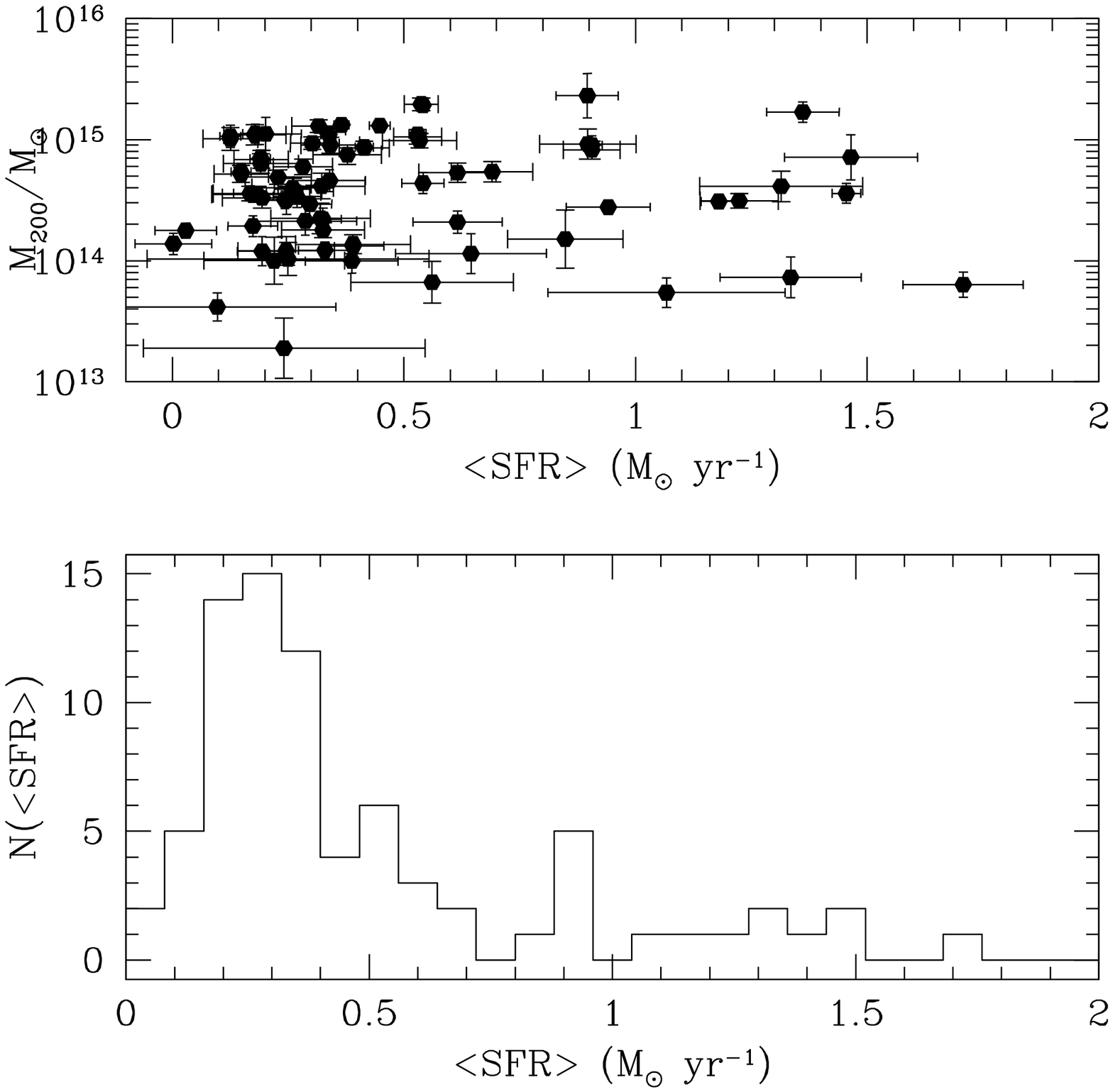}}
\end{minipage}
\end{center}
\caption{Upper panel: $M_{200}$ vs. the mean cluster SFR, $<SFR>$.
Lower panel: the distribution of $<SFR>$.}
\label{media}
\end{figure}

$\Sigma SFR$ is also significantly correlated with $\sigma_v$. The
best-fit parameters of the regression line are listed in Table
\ref{table1}. 
 
To check the robustness of our results we have re-analyzed the $\Sigma
SFR-M_{200}$ and $\Sigma SFR-\sigma_v$ relations by considering in
turn only the clusters with more than 20, 30 and 40 cluster
members. The correlations remain significant, and the values of the
best-fit parameters of the regression lines are consistent, within
errors, with those obtained when considering the whole cluster sample.

The correlation between $\Sigma SFR$ and $L_X$ is less well defined
than in the previous cases due to the larger scatter, but the
correlation is very significant also in this case (see Fig. \ref{lx}
and Table \ref{table1}). The large scatter is at least partially due
to the Abell X-ray-Underluminous (AXU) clusters (see
Fig.\ref{lx}). These systems are similar to the normal X-ray emitting
clusters in all their optically-derived properties but are generally
X-ray underluminous for their mass and optical luminosity (see paper
V for further details).

The significant correlations between $\Sigma SFR$ and the cluster
global quantities may not all be independent from one another.  In
fact, $\sigma_v$, $M_{200}$, and $L_X$ are all correlated quantities
(see, e.g., Paper III). They are also correlated with $N_{gal}$ (see
Paper VII), as it is $\Sigma SFR$ (see Fig.~\ref{ngal}, and Table
\ref{table1} -- note that the same de-projection correction applies to
both $\Sigma SFR$ and $N_{gal}$, so the relation between the two
quantities does not vary after applying this correction). We perform a
multiple regression analysis (e.g. Flury \& Riedwyl 1988; see also
Biviano et al. 1991 for another application of the method in an
astrophysical context) to try to understand which (if any) of these
correlations is the most fundamental one. We take $\Sigma SFR$ as the
dependent variable and consider $N_{gal}$, $\sigma_v$, $M_{200}$, and
$L_X$ as independent variables (regressors). We then adopt the method
of backward elimination (Flury \& Riedwyl 1988) in order to identify
the fundamental regressors for the dependent variable $\Sigma SFR$.
Namely, we compute the coefficient of determination, $R_p^2$, using
all $p$ regressors first, then eliminate each regressor one at a time
and look at the variation in $R_p^2$. The regressor giving the
smallest contribution to $R_p^2$ is eliminated, and we proceed until
only one regressor is left. When fundamental regressors are
eliminated, $R_p^2$ is substantially reduced. 

We find that the only fundamental regressor of $\Sigma SFR$ is
$N_{gal}$. I.e., $\Sigma SFR$ depends on $M_{200}$ (but also on
$\sigma_v$ and $L_X$) only because the more massive a cluster, the
larger its number of cluster galaxies and, proportionally, of
star-forming galaxies. 

Not only $N_{gal}$ is the fundamental regressor of $\Sigma SFR$, the
relation between the two quantities is linear. This may come as a
surprise if clusters of different richness contain different fraction
of star-forming galaxies within their virial radius. However, this is
not seen in our cluster sample (see Sect.~5). Since the $\Sigma SFR$
vs. $N_{gal}$ is linear (see Table \ref{table1}), the mean SFR of
cluster galaxies is constant (and equal to 0.47$\pm$0.13
$M_{\odot}/yr$). This is illustrated in Fig. \ref{media}, where we
show $M_{200}$ vs.  $<SFR>=\Sigma SFR/N_{gal}$ (upper panel) and the
$<SFR>$ distribution among our clusters. No significant relation is
found between $<SFR>$ and $M_{200}$ (nor in fact between $<SFR>$ and
either $\sigma_v$, or $L_X$). The scatter in the $<SFR>$ distribution
is at least partly due to the uncertainties in the incompleteness
correction factors (see Sect.~\ref{csfr}).

In lieu of normalizing $\Sigma SFR$ by the number of cluster members,
for the sake of comparison with other works in the literature, we also
normalize it by the cluster mass, $\Sigma SFR/M_{200}$. As expected
from the $\Sigma SFR$ vs. $M_{200}$ relation, there is no significant
trend of $\Sigma SFR/M_{200}$ with $M_{200}$, i.e. $\Sigma
SFR/M_{200}$ is constant\footnote{Note that we would have obtained a
significant anti-correlation of $\Sigma SFR/M_{200}$ with cluster
mass, had we not applied the de-projection correction.}.  Similarly,
there is no correlation between $\Sigma SFR/M_{200}$ and $L_X$.  The
evidence for a significant anti-correlation of $\Sigma SFR/M_{200}$
with $\sigma_v$ (see Table \ref{table1}) is somewhat surprising, given
that the slopes of the regression lines between $\Sigma SFR$ and
$M_{200}$, on the one side, and $\sigma_v$, on the other side, are
consistent with each-other ($2.18\pm0.23$ and $2.5\pm0.05$,
respectively, see Table \ref{table1} and paper III).  We note,
however, that the slope of the $\Sigma SFR/M_{200}-\sigma_v$ relation
is still consistent within $2\sigma$ with the value inferred from the
$\Sigma SFR-\sigma_v$ and $M_{200}-\sigma_v$ relations.

We conclude that the increase of $\Sigma SFR$ as a function of the
cluster mass is due to the proportionality between $\Sigma SFR$ and
$N_{gal}$ and that the mean SFR per galaxy or per unit mass is nearly
constant throughout our cluster sample, except perhaps for a residual
dependence on the cluster velocity dispersion.

\section{The total cluster stellar mass vs. the cluster global properties.}
We have performed a similar analysis as that described in the previous
Section using the total cluster stellar mass, $M_{\star}$, in lieu of
$\Sigma SFR$.  $M_{\star}$ is computed by summing up the stellar mass
of all the cluster spectroscopic members within $r_{200}$ and with
$M_r \leq -20.25$ (we use the median values of the stellar masses in
the B04 catalog). As for $\Sigma SFR$, we correct $M_{\star}$ for the
incompleteness and for the projection effects (see
Sect.~\ref{csfr}). As shown by the results listed in Table
\ref{table1}, the cluster $M_{\star}$ is proportional to $\Sigma
SFR$. As a consequence, the slopes of the relations of $N_{gal}$,
$M_{200}$, $\sigma_v$, and $L_X$ with $M_{\star}$ are all consistent
with those of the corresponding relations of these quantities with
$\Sigma SFR$. A multiple regression analysis shows that, also in this
case, the fundamental regressor of $M_{\star}$ is $N_{gal}$.

\begin{figure}
\begin{center}
\begin{minipage}{0.49\textwidth}
\resizebox{\hsize}{!}{\includegraphics{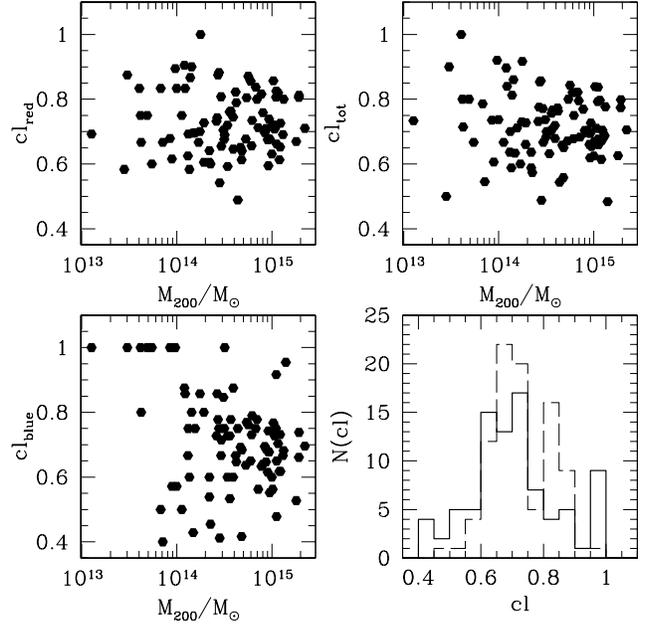}}
\end{minipage}
\end{center}
\caption{Spectroscopic completeness of the total (top-right panel),
red (top-left panel), and blue (bottom-left panel) cluster galaxy
populations as a function of the cluster mass. The bottom-right panel
shows the distributions of the whole (solid histogram) and blue
(dashed histogram) cluster galaxy populations.}
\label{complete}
\end{figure}

\begin{figure*}
\begin{center}
\begin{minipage}{0.31\textwidth}
\resizebox{\hsize}{!}{\includegraphics{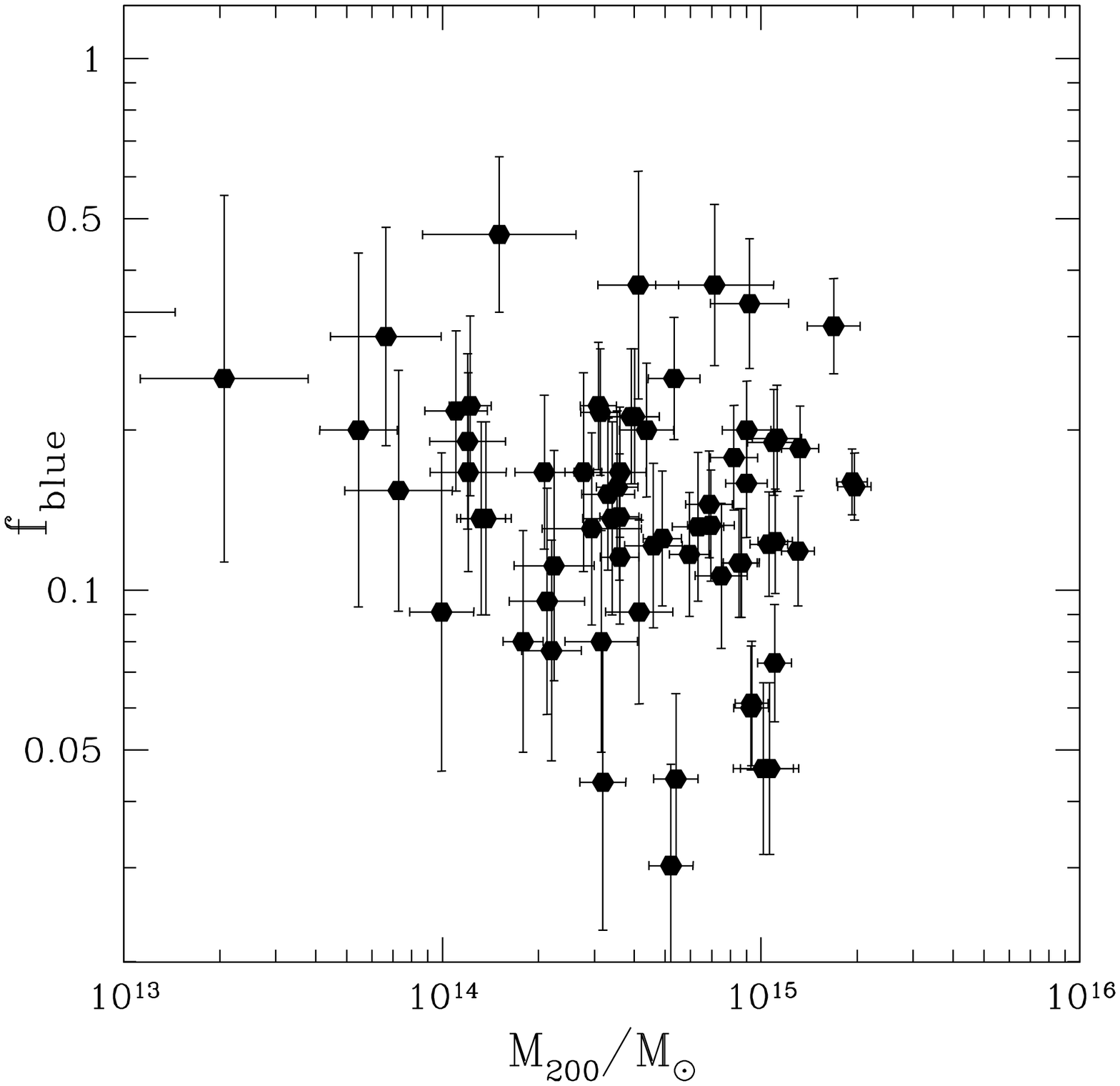}}
\end{minipage}
\begin{minipage}{0.31\textwidth}
\resizebox{\hsize}{!}{\includegraphics{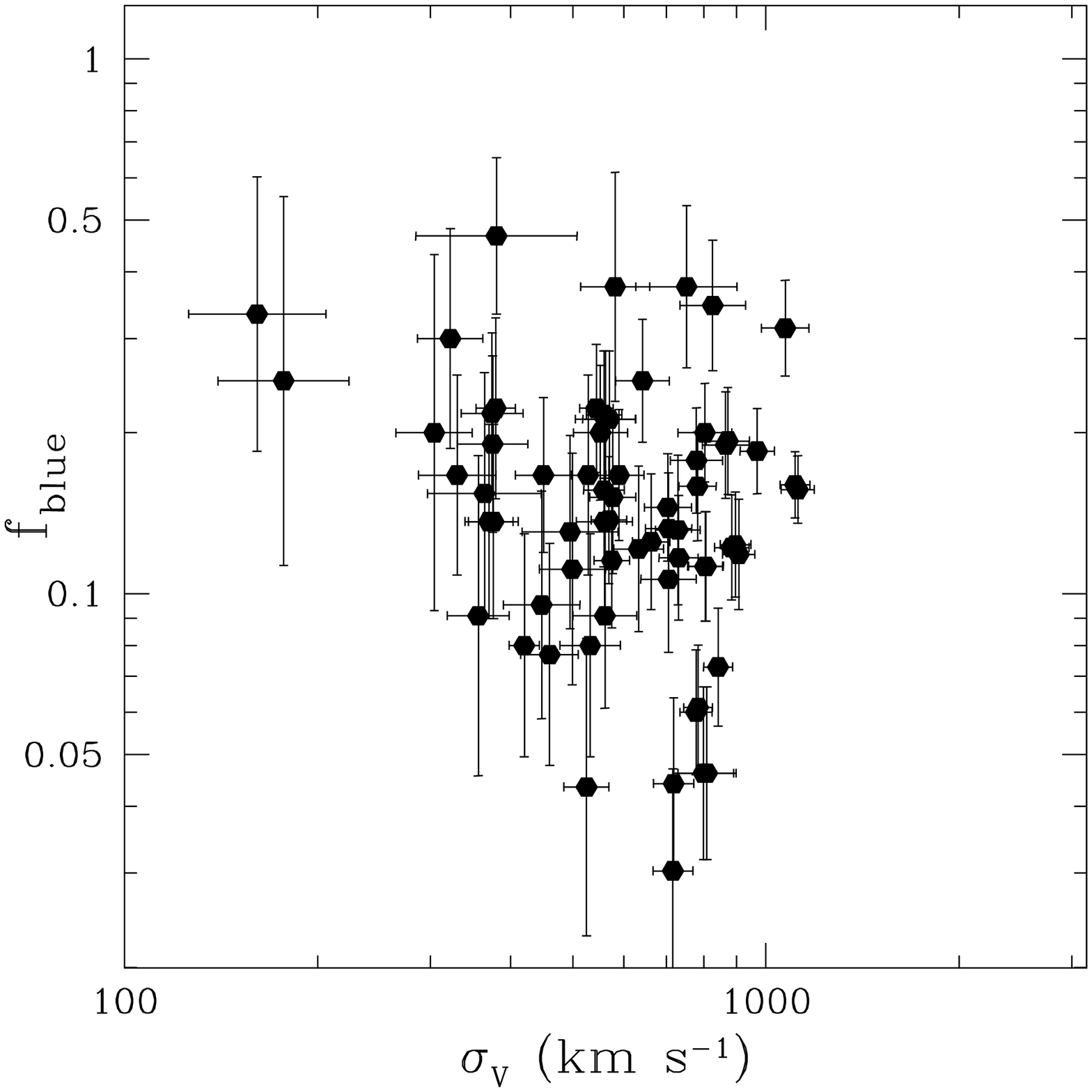}}
\end{minipage}
\begin{minipage}{0.31\textwidth}
\resizebox{\hsize}{!}{\includegraphics{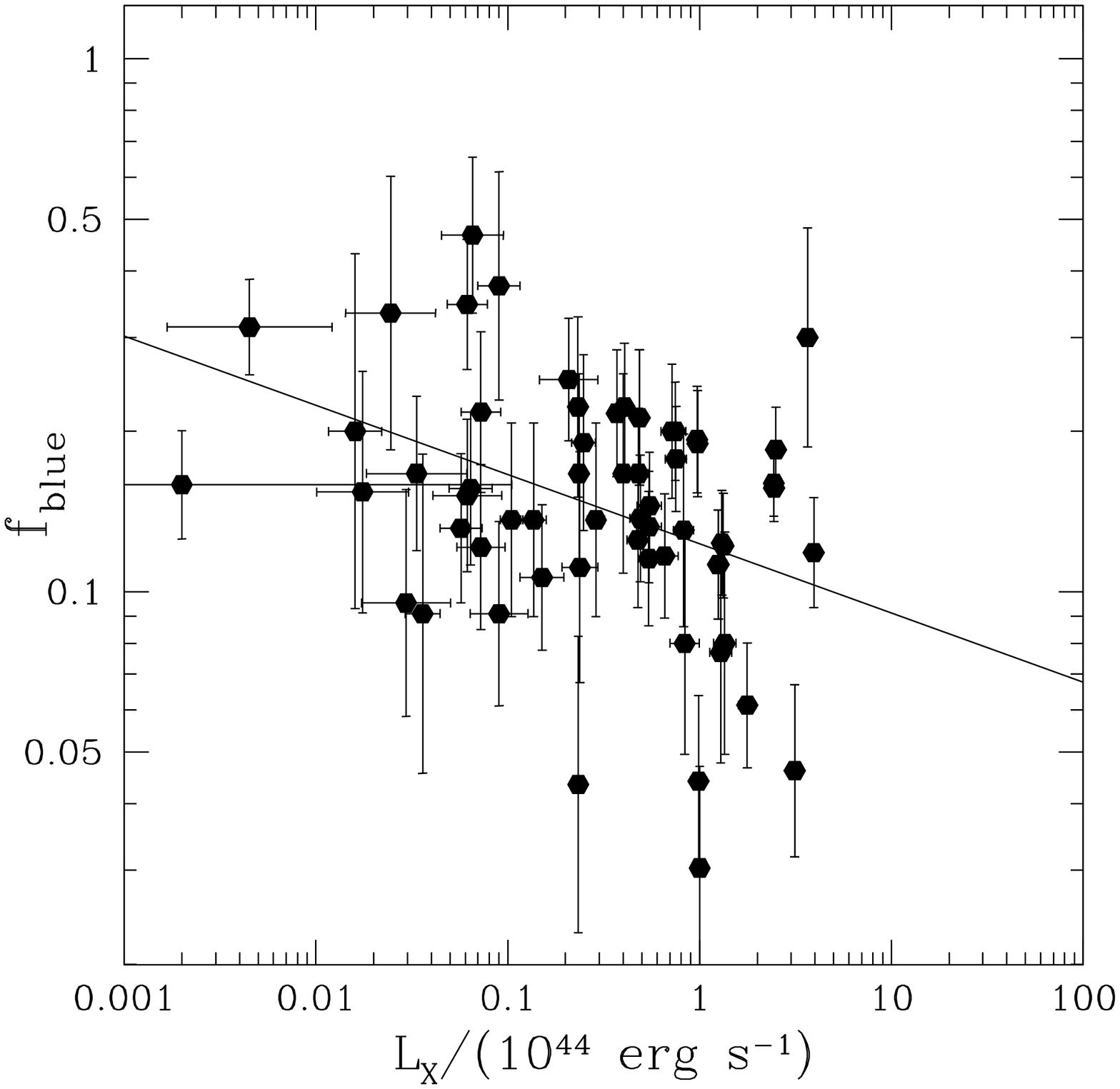}}
\end{minipage}
\end{center}
\caption{Relations between the fraction of blue cluster galaxies and
cluster global properties: $M_{200}$, $\sigma_v$, and $L_X$. The
best-fit regression lines are shown for the statistically significant
correlations only.}
\label{fb}
\end{figure*}

\begin{figure}
\begin{center}
\begin{minipage}{0.49\textwidth}
\resizebox{\hsize}{!}{\includegraphics{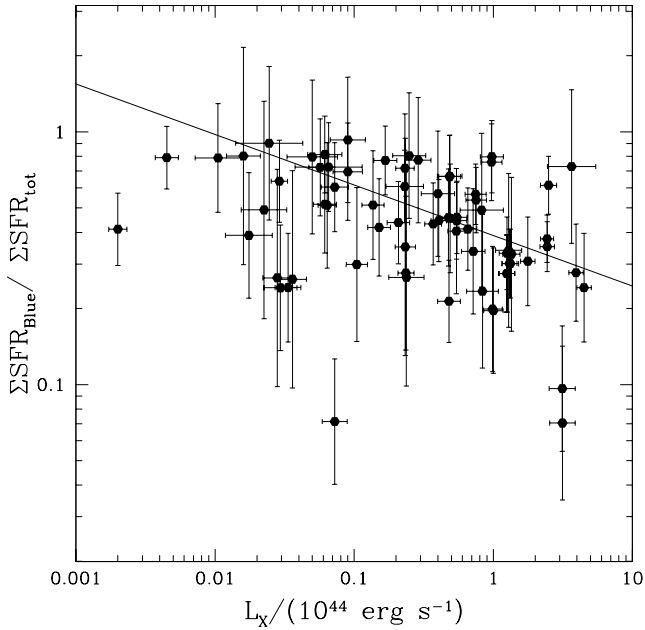}}
\end{minipage}
\end{center}
\caption{The relation between the fraction of $\Sigma SFR$ due 
to blue cluster galaxies and $L_X$.
The best-fit regression line is shown.}
\label{sfr_frac}
\end{figure}

\begin{figure}
\begin{center}
\begin{minipage}{0.49\textwidth}
\resizebox{\hsize}{!}{\includegraphics{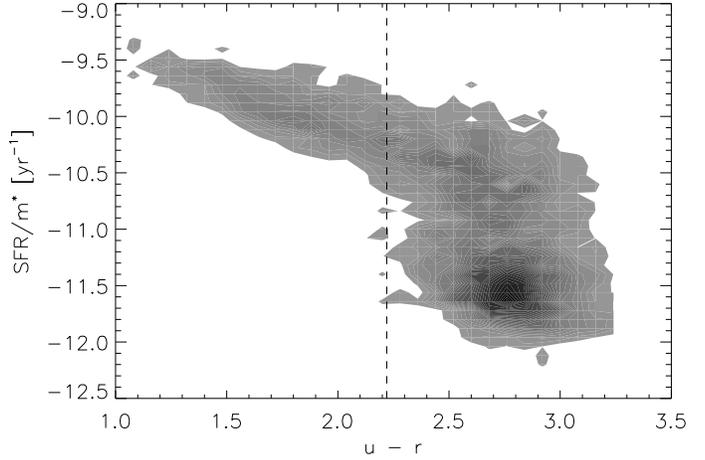}}
\end{minipage}
\end{center}
\caption{Star formation rate per unit of stellar mass versus the $u-r$
color for the cluster spectroscopic members.  The grey shading
intensity is proportional to the logarithm of the density 
of galaxies in the diagram.  The dashed line in the plot is the
color cut of Strateva et al. (2001) at $u-r=2.22$, used to separate
red from blue galaxies. Note that in addition to the usual populations
of star-forming blue galaxies and of no star-forming red galaxies,
there is a third population of red, star-forming red galaxies at
$SFR/m^* \ge 10^{-10.5} yr^{-1}$.}
\label{sfrh}
\end{figure}

\begin{figure}
\begin{center}
\begin{minipage}{0.49\textwidth}
\resizebox{\hsize}{!}{\includegraphics{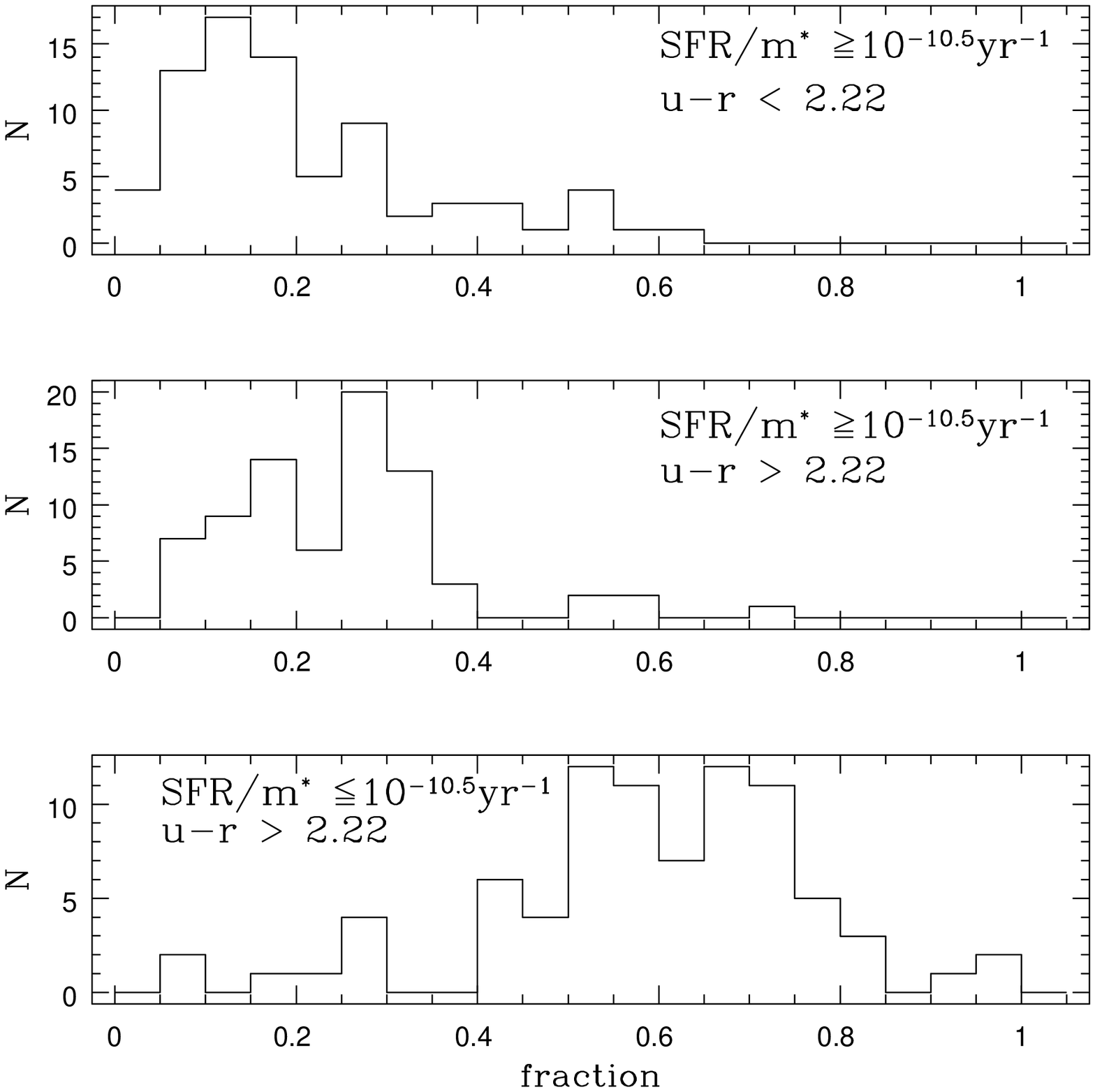}}
\end{minipage}
\end{center}
\caption{Distributions of the fractions of blue star-forming
galaxies (top panel), red star-forming galaxies (central panel)
and quiescent galaxies (bottom panel) in our cluster sample.}
\label{fraction_sfr}
\end{figure}

\section{The fractions of blue and star-forming galaxies vs. 
the cluster global properties.}
\label{s-frac}
We analyze the relations between the fractions of blue ($f_b$) and
star-forming ($f_{SF}$) galaxies in clusters with the cluster global
properties.  We define $f_b$ as the ratio between the number of
spectroscopic cluster members with $u-r<2.22$ (see Strateva et
al. 2001), and the number of all spectroscopic cluster members, within
$r_{200}$.  We do not need to apply an incompleteness correction here,
since we find that the blue and the whole cluster galaxy populations have
similar incompleteness levels for $r_{Petro} \leq -20.25$, within the
statistical uncertainties, as shown in Fig. \ref{complete} (the
incompleteness are estimated as in Sect.~\ref{csfr}, but taking into
account the color cuts). We do not apply the de-projection correction
either, since the de-projection correction factor for
the blue galaxies is very uncertain and in any case consistent with that
for the whole population.

The correlations of $f_b$ with $M_{200}$, $\sigma_v$, and $N_{gal}$
are not significant.  On the other hand, there is a significant
anti-correlation of $f_b$ with $L_X$ (see Table \ref{table1} and
Fig. \ref{fb}).

The $f_b$ vs. $L_X$ relation deserves a closer look. Another way of
looking at it is through the use of the fractional contribution of
blue galaxies to $\Sigma SFR$, $\Sigma SFR_{blue}/\Sigma
SFR_{tot}$. $\Sigma SFR_{blue}/\Sigma SFR_{tot}$ is anti-correlated
with $L_X$ (see Fig. \ref{sfr_frac}), and the slopes of the $f_b-L_X$
and $\Sigma SFR_{blue}/\Sigma SFR_{tot}-L_X$ relations are consistent
within the errors ( see Table \ref{table1}).

The color cut of Strateva et al. (2001) is used to separate blue from
red galaxies, but not all the star-forming galaxies are bluer than
$u-r=2.22$. Fig. \ref{sfrh} shows the $SFR/m^*$ in a sample of 2680
cluster galaxies versus the color $u-r$.  The dashed line in the plot
is the color cut of Strateva et al. (2001) at $u-r=2.22$. In addition
to the usual populations of star-forming blue galaxies and of no
star-forming (quiescent) red galaxies, there is a third population of
red, star-forming red galaxies at $SFR/m^* \ge 10^{-10.5}
yr^{-1}$. Hence, the color cut by itself does not distinguish between
star-forming and quiescent galaxies. For this we need a cut in
$SFR/m^*$, that we set at $SFR/m^*= 10^{-10.5} yr^{-1}$. We then
define $f_{SF}$ as the fractional number of galaxies with mass
normalized SFR above this limit. There is no significant correlation
of $f_{SF}$ with any cluster global quantity, $M_{200}$, $\sigma_v$,
$N_{gal}$, and $L_X$. Thus, while $f_b$ is anti-correlated with
$L_X$, $f_{SF}$ is not. This is due to the inclusion of the red
star-forming galaxies in the sample. In fact, the fraction of red
star-forming galaxies do not correlate with any of the global cluster
properties, not even $L_X$, and among the star-forming galaxies the
red ones outnumber the blue ones. This can be seen in
Fig. \ref{fraction_sfr}: the median fractions of blue star-forming,
red star-forming, and red quiescent galaxies are $0.15\pm0.05$,
$0.26\pm0.07$ and $0.63\pm0.05$.

\begin{figure}
\begin{center}
\begin{minipage}{0.49\textwidth}
\resizebox{\hsize}{!}{\includegraphics{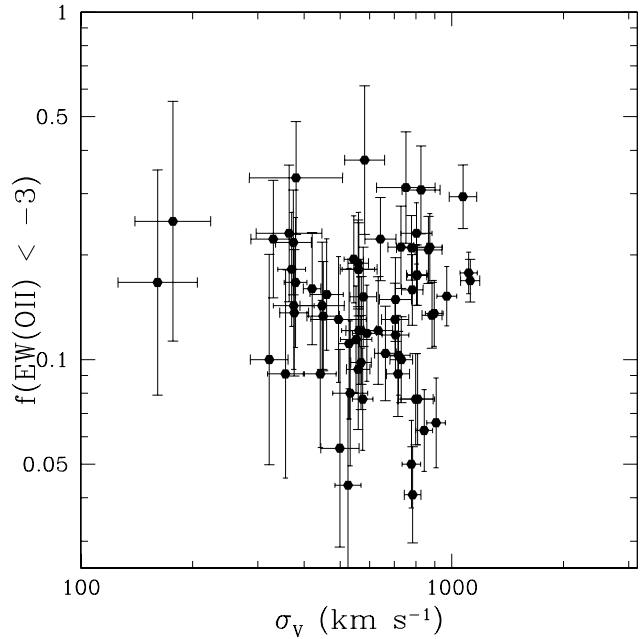}}
\end{minipage}
\end{center}
\caption{The fraction of ELGs vs. $\sigma_v$ in our cluster
sample. Note that AGNs and composite-spectra galaxies are
not included in our ELG galaxy sample. No significant
correlation exists.}
\label{oiisv}
\end{figure}

\section{Discussion}
Our results show that the cluster global properties ($M_{200}$,
$\sigma_v$, $L_X$) do not influence the SF properties of cluster
galaxies. While $\Sigma SFR$ {\em does} increase with increasing
cluster $M_{200}$, $\sigma_v$ and $L_X$, all these trends can be
totally explained as a richness effect, $\Sigma SFR \propto N_{gal}$.
The more galaxies in a cluster, the larger its mass, and the higher
its number of star-forming galaxies. Since the relation between
$\Sigma SFR$ and $N_{gal}$ is linear, the average cluster SFR is
essentially constant throughout our cluster sample.  Consistently, we
do not find any dependence of $f_{SF}$ with any cluster global
property. We do however find a residual correlation of the mass
normalized integrated SFR, $\Sigma SFR/M_{200}$ with $\sigma_v$, and a
significant anti-correlation of $f_b$ with $L_X$.

Also the total stellar mass, $M_{\star}$, depends linearly on
$N_{gal}$, i.e. the average stellar mass per cluster galaxy does not
depend on cluster properties. This is consistent with the universality
of the shape of the cluster luminosity function found in Paper IV. It
suggests that not only the average {\em current} star formation but
also the average {\em history} of star formation in clusters is
independent on the cluster properties .

How do our results compare with previous findings? The lack of
correlations we find between $f_b$ and $\sigma_v$, $M_{200}$, and
$N_{gal}$ confirm previous negative results by Goto (2005) and De
Propris et al. (2004) but disagree with the claimed trend of $f_b$
with cluster richness (Margoniner et al. 2001; Goto et al. 2003).  We
agree with Goto (2005) that there is no dependence of $\Sigma
SFR/M_{200}$ on $M_{200}$, but, at variance with his findings, we do
find a correlation between $\Sigma SFR/M_{200}$ and $\sigma_v$, as
well as between $\Sigma SFR$ and either $M_{200}$ or $\sigma_v$, in
broad agreement with the tentative correlations found by Homeier et
al. (2005).

Our results disagree with those of Finn et al. (2005), since, unlike
them, we do {\em not} find that the integrated SFR per cluster mass
decreases with increasing cluster mass. Our results disagree also with
those of Lin et al. (2003), since they find $M_{\star}/M_{500} \propto
M_{500}^{-0.26}$, while we find a linear relation between $M_{\star}$
and $M_{200}$, meaning that the fraction of mass in form of stars,
$M_{\star}/M_{200}$, is constant among different clusters. Remarkably,
however, our result would have been consistent with both Finn et al.'s
and Lin et al.'s had we also neglected to apply the mass-dependent
de-projection correction to $\Sigma SFR$ and $M_{\star}$ (see
Sect.~\ref{csfr}) as they did.

The anti-correlation we find between $f_b$ and $L_X$ is in disagreement
with previous claims of no correlations by Lea \& Henry (1988),
Fairley et al. (2002), and Wake et al. (2005). Such a correlation, as
well as the lack of correlation between $f_b$ and other cluster global
quantities, is however consistent with the result of Postman et
al. (2005). Postman et al. have recently shown that the fraction of
early-type galaxies in distant clusters increases with $L_X$, but does
not depend on either $\sigma_v$, or $T_X$. We actually checked that
the fraction of {\em red,} rather than blue, galaxies in our clusters
does show a relation with $L_X$ which is consistent (within 2
$\sigma$s) with the relation found by Postman et al. for their distant
cluster sample.

The lack of correlation we find between $f_{SF}$ and $L_X$ confirms
the results of Balogh et al. (2002), but the lack of correlation we
find between $f_{SF}$ and $\sigma_v$ is in disagreement with the
recent findings of P06. In their nearby cluster
sample, they find a decreasing fraction of ELGs with increasing
$\sigma_v$ for $\sigma_v \leq 500$ km~s$^{-1}$.

It is difficult to explore in detail the reasons for all the apparent
discrepancies among different results. One important issue is the
de-projection correction that we have introduced (see Sect.~\ref{csfr})
and that has not been applied before. Another important issue is the
limiting absolute magnitudes adopted in different studies. Yet another
relevant point could be the difference among different cluster
samples, since different samples span different redshift and mass
ranges, and none of the samples studied so far can be claimed to be a
volume-complete sample down to a given cluster mass limit.  Since
there is a significant overlap of the sample with P06, and we both use
data from the SDSS, we deem nevertheless worthwhile to investigate
further the reason why our results are in disagreement.

We first compared the values for the $\sigma_v$s of 22 clusters in
common. P06's and our values are very nicely correlated, and obey a
regression relation with a slope close to unity (although their values
are systematically higher than ours by $\sim 50$ km~s$^{-1}$). The
result discrepancy must origin in the different definition of the
fraction of star-forming galaxies. P06 define the star-forming
galaxies as those cluster members with a [OII] emission-line with
equivalent width (EW) smaller than $-3$~\AA. For the sake of
comparison we show in Fig.~\ref{oiisv} the relation between the
fraction of ELGs (with EW smaller than $-3$~\AA) and $\sigma_v$ in our
sample. At variance with P06 we do exclude AGNs and composite-spectra
galaxies from our sample. There is no significant correlation, no
trend is evident. Including AGNs in our sample we instead recover
the trend found by P06. Hence we conclude that the trend reported by
P06 is due to their including AGNs among the star-forming galaxies. We
will pursue the investigation of this topic in a forthcoming paper
(Popesso \& Biviano 2006).

Two relations that we find cannot be simply explained by the linear
relation between $\Sigma SFR$ and $N_{gal}$. These are the observed
decrease of $f_b$ with increasing $L_X$, and the observed decrease of
$\Sigma SFR/M_{200}$ with $\sigma_v$.  The fact that $f_b$ does not
correlate with $M_{200}$ excludes the possibility that the $f_b-L_X$
anti-correlation reflects a dependence of the fraction of blue
galaxies on cluster mass, as suggested by Postman et al. (2005). As a
matter of fact, $L_X$ is not a very good proxy for the cluster mass
(Reiprich \& B\"ohringer 2002; Paper III).  The anti-correlation
$f_b-L_X$ may be telling us more about the cluster and galaxy
formation processes than about the cluster evolution process. A
possible physical mechanism that could be responsible for this
anti-correlation is ram-pressure stripping (Gunn \& Gott 1972). The
ram -pressure force is proportional to $\rho_{ICM} \sigma_v^2$, where
$\rho_{ICM}$ is the density of the IC diffuse gas, and also $L_X$ is
proportional to $\rho_{ICM}^2$. If ram-pressure stripping is
indeed responsible for the $f_b$--$L_X$ anti-correlation, its strength
should depend on the clustercentric radius. Unfortunately our data are
not sufficient to test such a dependence.

The fact that the same anti-correlation is seen in high-$z$ clusters
(Postman et al.  2005) would argue for little evolution in the
properties of the IC gas out to $z \sim 1$, if ram-pressure stripping
is really the main process at work. Timescale is not a problem, since
ram-pressure stripping is a rapid process (Vollmer et
al. 2001). Because of the proportionality with $\sigma_v^2$,
ram-pressure stripping is also our best candidate for explaining the
anti-correlation of $\Sigma SFR/M_{200}$ with $\sigma_v$.

Although models of galaxy evolution in clusters tend to assign little
importance to the ram-pressure stripping mechanism (e.g. Okamoto \&
Nagashima 2003; Lanzoni et al. 2005), direct evidence for ongoing
ram-pressure stripping in cluster galaxies exist (e.g. Gavazzi et
al. 2003; Kenney et al. 2004). Ram-pressure is thought to induce gas
stripping from cluster galaxies, thereby reddening their
colors. However, the stripped gas can eventually fall back into the
aged galaxy, producing a short and mild burst of SF (Vollmer et
al. 2001; Fujita 2004), and this could explain why we observe an
anti-correlation between $f_b$ and $L_X$ but not between $f_{SF}$ and
$L_X$. 

As a matter of fact, $f_{SF}$ differs from $f_b$ because of the
presence of a red star-forming cluster galaxy population making up a
significant portion of the cluster star forming members, on average
25\% of the whole cluster galaxy population. The red colors
($u-r > 2.22$) of these galaxies suggest that they are dominated by an
old stellar population, unless there is a significant amount of dust
extinction.  The spectra of our red star-forming cluster galaxies are
similar to those of early-type spirals (Sa--Sb). Evidence for such a
population of red star-forming galaxies has already been found in
other studies (Demarco et al. 2005; Homeier et al. 2005; J\o rgensen
et al. 2005; Tran et al. 2005a, 2005b; Weinmann et
al. 2006). Their spectra are characterized as $k+a$ (Franx 1993) with
[OII] or H$\alpha$ (Miller et al. 2002) in emission. Their
morphologies are disklike (Tran et al. 2003), and their
concentrations are intermediate between those of the blue star-forming
and of those of the red and passive populations (Weinmann et
al. 2006).

We can interpret these red star-forming galaxies as objects in the
process of accomplishing their transformation from late- to early-type
galaxies. This transformation process may be identified by the
ram-pressure stripping because of the above mentioned correlations.
Another process able to induce bursts of SF in otherwise quiescent
galaxies is the merger of two quiescent galaxies. While the process
could occur in distant, low-$\sigma_v$ clusters (Tran et al. 2005b),
it is very unlikely to be effective in nearby ones (e.g. Mamon 1996).
Fast encounters between galaxies in clusters rather produce the
'harassment' mechanism described by Moore et al. (1996, 1998).

Recently, these red star-forming galaxies have also been found outside
clusters. According to Franzetti et al. (2006), $\sim 35$--40\% of all
the red field galaxies have ongoing SF. This fraction is comparable,
if not higher, than the fraction we observe in our sample of nearby
clusters, and suggest that we do not actually need a cluster-related
phenomenon to explain the presence of red star-forming
galaxies. Perhaps these galaxies are simply more dusty than the blue
star-forming galaxies (e.g. Tran et al. 2005a).  Red-sequence
mid-infrared emitters, with significant levels of inferred SF, have
indeed already been detected in some clusters (Miller et al. 2002;
Biviano et al. 2004; Coia et al.  2005).

In conclusion, we feel that a more detailed analysis of the morphology
of the red star-forming systems and a careful study of their
properties within and outside the cluster environment, are mandatory
for understanding their nature.

\section{Conclusion}
\label{s-conc}
We have analyzed the relationships between SF in cluster galaxies and
global cluster properties, such as cluster $M_{200}$, $\sigma_v$,
$L_X$, and $N_{gal}$.  For our analysis we have used a sample of 79
nearby clusters extracted from the RASS-SDSS galaxy cluster catalogue
of Paper III and Paper V. Galaxy $SFR$s and stellar masses are taken
from the catalog of Brinchmann et al. (2004), which is based on SDSS
spectra. We only consider galaxies with $M_r \leq -20.25$ in our
analysis, and exclude AGNs and composite-spectra galaxies.

All the cluster quantities considered are corrected for incompleteness,
when needed, and for projection effects. The de-projection correction
is of particular importance in our analysis, since we find that it
depends on the cluster mass.

$\Sigma SFR$ is correlated with all the cluster global quantities
mentioned above.  By performing a multiple regression analysis that
the main correlation is that between $\Sigma SFR$ and $N_{gal}$. Since
this relation is linear the average SFR of cluster
galaxies is the same in different clusters, and is unaffected by
either the cluster mass, or its velocity dispersion, or its X-ray
luminosity.  We come to essentially the same conclusion when
$M_{\star}$ is considered in lieu of $\Sigma SFR$. If instead of
normalizing $\Sigma SFR$ with $N_{gal}$ we normalize it with
$M_{200}$, we still find $\Sigma SFR/M_{200}$ does not depend on any
cluster global property, except $\sigma_v$, which we suggest could be
evidence of the effect of ram-pressure stripping on the cluster galaxy
properties. 

Ram-pressure could also be the mechanism able to explain the observed
anti-correlation of $f_b$ with $L_X$, since $f_b$ is {\em not}
correlated with either $M_{200}$ or with $\sigma_v$. On the other
hand, the fact that we do not observe any correlation between $L_X$
and $f_{SF}$ is due to the presence of a dominant fraction of {\em
red} star-forming galaxies. They could also be the result of the
ram-pressure mechanism, or, in alternative, they could be star-forming
galaxies with an anomalous amount of dust.

If global cluster properties affect the star-forming
properties of cluster galaxies, their effect is rather marginal,
except perhaps on galaxy colors, which seem to be influenced by
the presence of the IC diffuse gas.

\vspace{2cm}

We thank the anonymous referee for useful suggestions which helped
us improving the quality of this paper.
Funding for the creation and distribution of the SDSS Archive has been
provided by the Alfred P.  Sloan Foundation, the Participating
Institutions, the National Aeronautics and Space Administration, the
National Science Foundation, the U.S.  Department of Energy, the
Japanese Monbukagakusho, and the Max Planck Society. The SDSS Web site
is http://www.sdss.org/. The SDSS is managed by the Astrophysical
Research Consortium (ARC) for the Participating Institutions.  The
Participating Institutions are The University of Chicago, Fermilab,
the Institute for Advanced Study, the Japan Participation Group, The
Johns Hopkins University, Los Alamos National Laboratory, the
Max-Planck-Institute for Astronomy (MPIA), the Max-Planck-Institute
for Astrophysics (MPA), New Mexico State University, University of
Pittsburgh, Princeton University, the United States Naval Observatory,
and the University of Washington.

\end{document}